# Continuous diffusion model for concentration dependence of nitroxide EPR parameters in normal and supercooled water

Dalibor Merunka[a,*] and Miroslav Peric[b]

[a]Division of Physical Chemistry, Ruđer Bošković Institute, Bijenička cesta 54, HR-10000 Zagreb, Croatia, E-mail: merunka@irb.hr

[b]Department of Physics and Astronomy and The Center for Supramolecular Studies, California State University at Northridge, Northridge, California 91330, United States, E-mail: miroslav.peric@csun.edu

January 5, 2017

[*]Corresponding author

**Abstract**

We measured electron paramagnetic resonance (EPR) spectra of $^{14}$N- and $^{15}$N-labeled perdeuterated TEMPONE radicals in normal and supercooled water at various radical concentrations. By fitting the EPR spectra to spectral shape functions based on the modified Bloch equations, we obtained concentration dependences of EPR parameters of radicals at each measured temperature. From concentration dependences of the EPR parameters quantifying spin dephasing, coherence transfer, and hyperfine splitting, we determined linear concentration coefficients, whose values depend on the relative motion of radicals due to modulation of the Heisenberg spin exchange (HSE) and dipole-dipole (DD) interactions between them. We applied the continuous diffusion model for relative motion of radicals and we evaluated the diffusion coefficients of radicals from the concentration coefficients using the standard relations and the relations derived from kinetic equations for the spin evolution of interacting radical pair. It was found that the latter equations lead to the better agreement between the diffusion coefficients calculated from different concentration coefficients. The calculated diffusion coefficients of $^{14}$N- and $^{15}$N-labeled radicals show similar values, which is an expected result that supports the presented method. Upon lowering the temperature into the supercooled state, the calculated diffusion coefficients decrease slower than is predicted by the Stokes-Einstein relation and slower than the rotational diffusion coefficient. Similar effects were detected in NMR studies of the rotational and translational motion of water molecules in supercooled water.

# 1. Introduction

Electron paramagnetic resonance (EPR) spectroscopy, a sensitive technique for detecting radicals in materials, can provide information about the translational motion of radicals in a liquid solution. The relative motion of radicals modulates the Heisenberg spin exchange (HSE) and dipole-dipole (DD) interactions between them, affecting the shape of the EPR spectrum [1,2]. The HSE interaction as a tool for finding the collision rates of radicals and their diffusion coefficient has a long time history [1]. The HSE method is based on the determination of the spin exchange frequency, which is proportional to the radical concentration and the diffusion coefficient of radical. This method has been applied to study the diffusion of radicals in various systems including liquids, liquid crystals, biological systems, porous hosts, etc. [1,3-8].

Traditionally, the spin exchange frequency is obtained from the concentration induced broadening of EPR lines, using the fact that the HSE interaction induces extra spin dephasing of the radical's magnetization, which in turn broadens the EPR lines [1,3-8]. The line-broadening method was found to be effective at high values of the diffusion coefficient of the radical, where the contribution of HSE interaction to line broadening dominates over that of DD interaction, which is practically averaged out. As the diffusion coefficient decreases, the line broadening due to HSE interaction decreases and that due to DD interaction increases, which makes the line broadening insensitive to changes in values of the diffusion coefficient. Since biologically important systems are often viscous enough to be in this regime [3], there is a need to study the concentration dependence of other EPR parameters, as well as to separate the effects of HSE and DD interactions on measured concentration dependences. The usual way of separation in the line-broadening method relies on assumptions that total broadening is a sum of the HSE and DD contributions that depend, respectively, directly and inversely on the diffusion coefficient whose temperature dependence follows Arrhenius behavior [5-8].

A more advanced approach to analysis of experimental EPR spectra affected by HSE interaction has been developed in the series of articles, where Ref. [9] is the first and Ref. [10] is the most recent part. In the initial study [9], nonlinear least-squares fitting was for the first time employed to fit spin exchange broadened EPR spectra. It was shown that the EPR line shape of a nitroxide free radical undergoing HSE in the slow exchange limit could be fit to the sum of first derivative Lorentzian absorption and dispersion lines, which is in agreement with theoretical result of Molin at al. [1]. The fitting of EPR spectra provides two additional parameters for extraction of the spin exchange frequency: (i) the ratio between amplitudes of dispersion and absorption components, and (ii) the shifts of the resonance fields of the outer absorption lines. The first parameter was found to be a linear function of spin exchange frequency [9], in accordance with theory [1]. The line shifts, or in other words, the absorption hyperfine spacing was found to have one linear and one quadratic term in the spin exchange frequency [11]. This finding is also in agreement with theoretical considerations of the HSE induced line shifts [12]. Additionally, a more elaborate separation method of the HSE and DD contributions was developed in this approach and applied on the stable nitroxide radical, perdeuterated 2,2,6,6-tetramethyl-4-oxopiperidine-1-oxyl ($^{14}$N-pDTEMPONE) in squalene, a viscous alkane [13]. This method uses Salikhov's theoretical analysis of the effects of HSE and DD interactions on EPR spectra [2], without making any assumption on the temperature dependence of diffusion coefficient. The separation is based on approximations applied to the sums of spectral densities of correlation functions for DD interaction, which define the DD contributions to spin coherence transfer and spin dephasing [13]. In the most recent part of the series [10], it was shown computationally that the sum of Lorentzian absorption and dispersion line shapes can be used to fit the EPR spectra of $^{15}$N- and $^{14}$N-labeled nitroxide radicals undergoing HSE even after the spectra have coalesced and have started to narrow.

Recently, Salikhov has shown theoretically that the EPR spectrum at any coherence transfer rate can be expressed as a sum of Lorentzian absorption and dispersion curves [14].

The above approach was also applied to study the diffusion of $^{14}$N-pDTEMPONE in the normal and supercooled states of water [15], which was the first EPR study of translational diffusion in supercooled water. The values of the diffusion coefficient derived from the separated HSE and DD contributions to the spin dephasing were found to be close to each other and their temperature dependences were found to follow the hydrodynamic behavior. On the other hand, the diffusion coefficient derived from the concentration dependence of hyperfine splitting was found not to follow the hydrodynamic behavior.

To further investigate this unusual behavior, we performed EPR measurements of both $^{14}$N-pDTEMPONE and pDTEMPONE labeled with $^{15}$N ($^{15}$N-pDTEMPONE) in normal and supercooled water. The choice of these two radicals could be useful for testing the present approach and its further development, because the EPR spectra of these radicals sharply differ, while their size and diffusion coefficient should be the same. Here, we solved the modified Bloch equations for both radicals in the presence of HSE and DD interactions. The obtained solutions were employed to fit experimental EPR spectra and extract the EPR spectral parameters for both radicals. This fitting approach, which is referred as the original function fitting method, was compared to the fitting approach based on the sum of absorption and dispersion lines [9,10,14], which is referred as the sum function fitting method. Using concentration dependences of the fitted values of EPR parameters that quantify spin dephasing, spin coherence transfer, and hyperfine splitting, we applied linear fitting and determined corresponding linear concentration coefficients. In the theoretical part of the study, we applied the continuous diffusion model for the relative motion of radicals. We evaluated the diffusion coefficients of radicals from the standard relations for concentration coefficients and the relations derived by iterative solving of the kinetic equations for the spin evolution of interacting radical pair. It was found that the latter equations predict the normal hydrodynamic behavior of the diffusion coefficients derived from the hyperfine splitting coefficients, as opposed to the standard relations. Additionally, these equations predict similar values of the diffusion coefficients calculated from all three concentration coefficients for both radicals. The temperature dependences of the calculated diffusion coefficients of radicals were compared to the Stokes-Einstein relation and the temperature dependence of the rotational diffusion coefficient of $^{14}$N-pDTEMPONE. The obtained results were compared to the temperature dependence of the rotational and translational motion of water molecules in supercooled water.

## 2. Materials and methods

The spin probes $^{15}$N-pDTEMPONE (Lot# Z447P4, 98 atom % D, 99 atom % $^{15}$N) and $^{14}$N-pDTEMPONE (Lot# P607P7, 99 atom % D) were purchased from CDN Isotopes and used as received. The stock solutions of 50 mM $^{15}$N-pDTEMPONE and 40 mM $^{14}$N-pDTEMPONE were prepared by weight in Milli-Q water. The solutions were diluted to 25 concentrations of $^{15}$N-pDTEMPONE (from 2 mM to 50 mM in steps of 2 mM) and 20 concentrations of $^{14}$N-pDTEMPONE (from 2 mM to 40 mM in steps of 2 mM). The samples were drawn into 5-μL capillaries (radius ≈ 150 μm) and sealed at both ends by an open flame. EPR spectra were recorded with a Varian E-109 X-band spectrometer upgraded with a Bruker microwave bridge and a Bruker high-Q cavity. The sample temperature was controlled by a Bruker variable temperature unit, and it was held stable within ±0.2 K. Temperature was measured with a thermocouple using an Omega temperature indicator. The thermocouple tip was always positioned at the top of the active region of the EPR cavity, to avoid reducing the cavity quality factor. All samples were measured in steps of 2 K in a temperature range from 253 to 283 K and in steps of 4 K in a temperature range from 283 to 303 K. The EPR spectra for both probes were acquired employing a sweep time of 20 s, microwave power of 0.5 mW, time constant of 16 ms, and modulation amplitude of 0.1 G. The sweep widths of 40 and 50 G were employed for $^{15}$N- and $^{14}$N-pDTEMPONE, respectively.

## 3. Fitting procedure and concentration coefficients

The experimental EPR spectrum $S(B)$ is the first derivative of absorption EPR signal $R(B)$ with respect to the applied magnetic field $B$, i.e., $S(B) = dR(B)/dB$. The modified Bloch equations for radicals with hyperfine structure in the presence of HSE can be solved in a closed mathematical form [1]. This was done for the equations of $^{15}$N- and $^{14}$N-labeled nitroxide radicals interacting by both HSE and DD interactions (Appendix A). The solution for absorption spectra of both radicals, referred as the original function (OF), has the form:

$$R(B) = J_0 \, \text{Re}\left[\frac{G(B)}{1 - \Lambda G(B)}\right]; \; G(B) = \sum_{k=1}^{2I+1} \frac{1}{z_k + \Lambda + i(B - B_0)}. \quad (1a)$$

Here, $J_0$ is a constant, $\Lambda$ is a coherence transfer rate in magnetic-field units, $B_0$ is central field position of the spectrum, and $2I+1$ is the number of hyperfine lines, where the spin of nitrogen nucleus have values $I = 1/2$ for $^{15}$N and $I = 1$ for $^{14}$N. The parameters $z_k$ are

$$z_1 = \Gamma_1 - iA/2; \; z_2 = \Gamma_2 + iA/2 \quad (1b)$$

for $^{15}$N-labeled radical, and

$$z_1 = \Gamma_1 - i(A + \Delta A/3); \; z_2 = \Gamma_2 + 2i\Delta A/3; \; z_3 = \Gamma_3 + i(A - \Delta A/3) \quad (1c)$$

for $^{14}$N-labeled radical. In these relations, $\Gamma_k$ is the spin dephasing rate or linewidth of the hyperfine line $k$, $A$ is the hyperfine coupling constant, and $\Delta A$ is the relative second-order hyperfine shift. In the OF fitting method, the first derivative of line shape function $R(B)$ defined by the expressions (1) is used to fit the experimental EPR spectra. The experimental spectra were transferred to a personal computer and fitting was performed using nonlinear regression command in the program package Mathematica. The experimental spectra for 40 mM solutions at 295 K are shown in Fig. 1, together with the fitting curves and residuals. The fits are quite good, as can be seen from the residuals, and the fitted OF parameters are quite accurately obtained (Table 1).

Fig. 1. Experimental EPR spectra (black lines), OF fitting curves (red lines), and residuals (green lines) of 40 mM aqueous solution of (a) $^{15}$N-pDTEMPONE and (b) $^{14}$N-pDTEMPONE at 295 K.

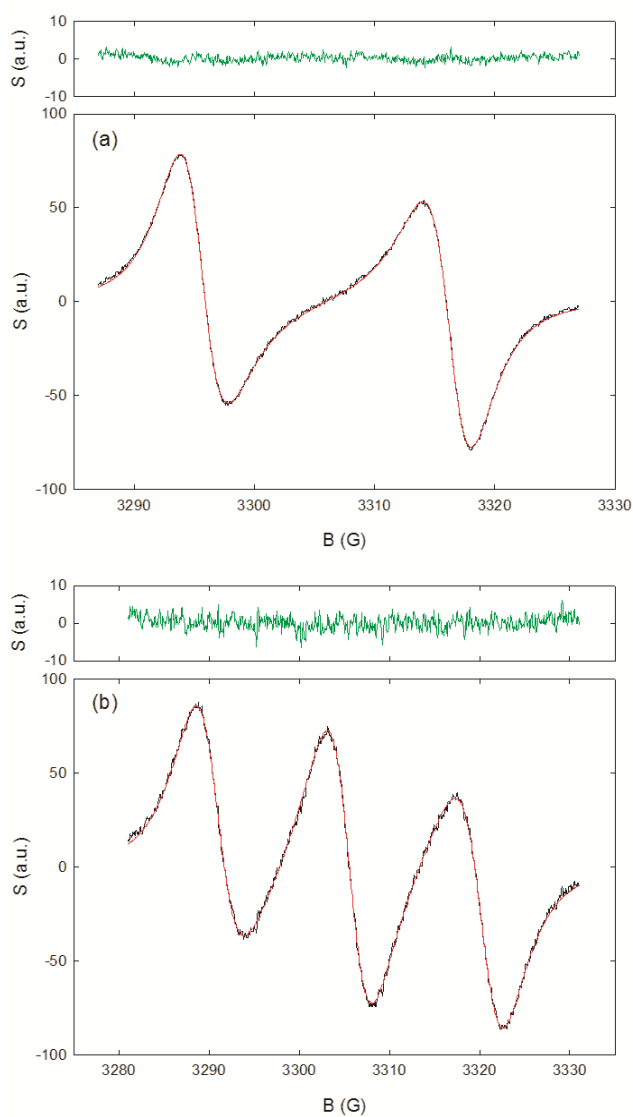

**Table 1.** The values and standard errors (in parentheses) of the original (OF) and sum function (SF) parameters for 40 mM aqueous solution of $^{15}$N-and $^{14}$N-pDTEMPONE at 295 K. The fitted OF and SF parameters are shown together with the OF parameters calculated from fitted SF parameters using exact and simplified relations.

| $^{15}$N OF | fit | exact | simplified | $^{15}$N SF | fit |
|---|---|---|---|---|---|
| $B_0$(G) | 3305.943(2) | 3305.976(3) | | $B_0$(G) | 3305.976(3) |
| $A$(G) | 21.90(2) | 21.90(1) | 21.90(1) | $A'$(G) | 21.063(6) |
| $\Gamma_1$(G) | 3.462(4) | 3.458(4) | | $\Gamma'_1$(G) | 3.458(4) |
| $\Gamma_2$(G) | 3.442(4) | 3.446(4) | | $\Gamma'_2$(G) | 3.446(4) |
| $\Lambda$(G) | 3.00(2) | 3.00(2) | 3.00(2) | $f_1^R$ (a.u.) | 1164(3) |
| $J_0$(a.u.) | 1173(2) | 1173(2) | | $f_2^R$ (a.u.) | 1183(3) |
| | | | | $f_1^I$ (a.u.) | −356(3) |
| | | | | $f_2^I$ (a.u.) | 312(3) |
| $\Gamma$(G) | 3.452(3) | 3.452(3) | 3.452(3) | $\Gamma'$(G) | 3.452(3) |

| $^{14}$N OF | fit | exact | simplified | $^{14}$N SF | fit |
|---|---|---|---|---|---|
| $B_0$(G) | 3305.572(3) | 3305.588(7) | | $B_0$(G) | 3305.588(7) |
| $A$(G) | 15.54(1) | 15.58(1) | 15.58(1) | $A'$(G) | 15.138(9) |
| $\Delta A$(G) | 0.012(5) | 0.00(2) | | $\Delta A'$(G) | 0.00(1) |
| $\Gamma_1$(G) | 4.709(9) | 4.68(1) | | $\Gamma'_1$(G) | 4.73(1) |
| $\Gamma_2$(G) | 4.618(7) | 4.59(2) | | $\Gamma'_2$(G) | 4.50(1) |
| $\Gamma_3$(G) | 4.707(9) | 4.75(1) | | $\Gamma'_3$(G) | 4.80(1) |
| $\Lambda$(G) | 2.08(2) | 2.14(2) | 2.13(2) | $f_1^R$ (a.u.) | 2050(10) |
| $J_0$(a.u.) | 2218(8) | 2209(6) | | $f_2^R$ (a.u.) | 2450(10) |
| | | | | $f_3^R$ (a.u.) | 2120(10) |
| | | | | $f_1^I$ (a.u.) | −930(10) |
| | | | | $f_2^I$ (a.u.) | −30(10) |
| | | | | $f_3^I$ (a.u.) | 930(10) |
| $\Gamma$(G) | 4.678(5) | 4.676(7) | 4.676(7) | $\Gamma'$(G) | 4.676(7) |

The OF fitting method is compared the sum function (SF) fitting method [9,10,14], which uses the fact that the absorption EPR signal (1) can be written as a sum of Lorentzian absorption and dispersion lines [14]:

$$R(B) = \text{Re}\left[\sum_k \frac{f_k}{z'_k + i(B - B_0)}\right]. \qquad (2a)$$

Here, $f_k = f_k^R + i f_k^I$ are the complex amplitudes of hyperfine lines, whose real and imaginary parts correspond to the intensities of absorption and dispersion components, respectively, while the parameters $z'_j$ for $^{15}$N-labeled radical are given by:

$$z'_1 = \Gamma'_1 - iA'/2; \quad z'_2 = \Gamma'_2 + iA'/2 \tag{2b}$$

and those for $^{14}$N-labeled radical are given by:

$$z'_1 = \Gamma'_1 - i(A' + \Delta A'/3); \quad z'_2 = \Gamma'_2 + 2i\Delta A'/3; \quad z'_3 = \Gamma'_3 + i(A' - \Delta A'/3). \tag{2c}$$

Here, $\Gamma'_k$, $A'$, and $\Delta A'$ are the linewidths, hyperfine splitting, and second-order hyperfine shift of absorption lines. The SF fitting method based on Eqs. (2) was found to be more stable and much less time consuming than the OF method, providing the same quality of fits. However, the fitted SF parameters are somewhat different from the corresponding OF parameters and the coherence transfer parameter $\Lambda$ should be calculated (Table 1). In order to calculate the OF parameters from the fitted SF parameters, we employ the exact relations between those parameters given in Appendix A [Eqs. (A6)]. The fitted OF parameters and those calculated by exact relations exhibit a fair agreement (Table 1), which supports consistency of both fitting methods. However, the evaluating of values and errors of OF parameters using exact relations was found to be somewhat inconvenient since they include equations with complex numbers. Therefore, we proposed simplified relations for three important OF parameters: the average linewidth of hyperfine lines $\Gamma$, the coherence transfer rate $\Lambda$, and the hyperfine coupling constant $A$. The simplified relations are given by (see Appendix A):

$$\Gamma = \Gamma'; \quad \Lambda \cong \frac{A'}{2} \frac{f_2^I - f_1^I}{f_1^R + f_2^R}; \quad A \cong \sqrt{A'^2 + 4\Lambda^2} \tag{3a}$$

for $^{15}$N-labeled radical, and by:

$$\Gamma = \Gamma'; \quad \Lambda \cong \frac{A'}{2} \frac{f_3^I - f_1^I}{f_1^R + f_2^R + f_3^R}; \quad A \cong \sqrt{A'^2 + 3\Lambda^2} \tag{3b}$$

for $^{14}$N-labeled radical. These relations are more practical than exact relations to calculate the values and errors of OF parameters without losing the accuracy of obtained results (Table 1).

**Fig. 2.** Concentration dependences of the spin dephasing rates $\Gamma_i$, the coherence transfer rate $\Lambda$, and the nitrogen hyperfine splitting $A$ for **(a)** $^{15}$N-pDTEMPONE and **(b)** $^{14}$N-pDTEMPONE in water at 295 K. Errors of EPR parameters are smaller than symbols. Lines denote linear fits of concentration dependences.

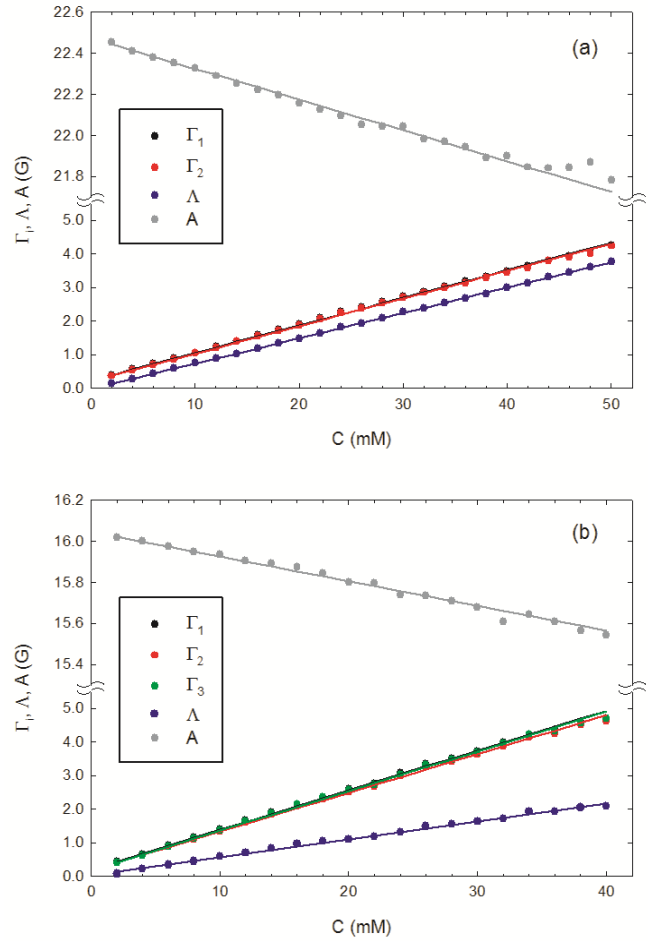

The HSE and DD interactions between radicals affect the dependences of EPR parameters $\Gamma_k$, $\Lambda$, and $A$ on radical concentration [2]. When the radical concentration $C$ is small enough, the parameters depend linearly on $C$, i.e.:

$$\Gamma_k = \Gamma_{k0} + W_{kj}C; \quad \Lambda = \Lambda_0 + V_j C; \quad A = A_0 - B_j C, \qquad (4)$$

where $W_{kj}$, $V_j$, and $B_j$ are the linear concentration coefficients of the $k$-th linewidth, the coherence transfer rate, and the hyperfine splitting, respectively. The type of radical is denoted by the index $j = 2I$, having values 1 and 2 for $^{15}$N- and $^{14}$N-pDTEMPONE, respectively [2]. The fitted OF parameters $\Gamma_k$, $\Lambda$, and $A$ show a linear dependence on $C$ in the measured concentration range indeed (Fig. 2). Therefore, the concentration coefficients of these OF parameters can be evaluated as the slopes of linear fits. The linear fits (Fig. 2) and the corresponding concentration coefficients (Table 2) are obtained by the weighted linear regression method, where the weights are inverse squares of the standard errors from OF fitting procedure (Table 1). We calculated the average concentration coefficients of hyperfine lines $W_j$, given by $W_1 = (W_{11} + W_{12})/2$ for $^{15}$N-pDTEMPONE and $W_2 = (W_{12} + W_{22} + W_{32})/3$ for $^{14}$N-pDTEMPONE. These average coefficients coincide with the concentration coefficients $W_{\Gamma j}$ of the average linewidths $\Gamma$ (Table 2).

**Table 2.** Values and standard errors of linear concentration coefficients of original function (OF) parameters for aqueous solution of $^{15}$N- and $^{14}$N-pDTEMPONE at 295 K. The coefficients are calculated for the fitted OF parameters and those calculated by exact and simplified relations from SF parameters.

| $^{15}$N | OF fit | OF exact | OF simplified |
|---|---|---|---|
| $W_{11}$(G/mM) | 0.0821(5) | 0.0821(5) | |
| $W_{21}$(G/mM) | 0.0820(5) | 0.0823(5) | |
| $W_1$(G/mM) | 0.0821(4) | 0.0822(4) | |
| $W_{\Gamma 1}$(G/mM) | 0.0821(5) | 0.0822(5) | 0.0822(5) |
| $V_1$(G/mM) | 0.0755(3) | 0.0752(3) | 0.0752(3) |
| $B_1$(G/mM) | 0.0150(3) | 0.0151(3) | 0.0151(3) |

| $^{14}$N | OF fit | OF exact | OF simplified |
|---|---|---|---|
| $W_{12}$(G/mM) | 0.1181(8) | 0.1178(8) | |
| $W_{22}$(G/mM) | 0.1156(8) | 0.1155(8) | |
| $W_{32}$(G/mM) | 0.1184(8) | 0.1190(8) | |
| $W_2$(G/mM) | 0.1174(4) | 0.1175(5) | |
| $W_{\Gamma 2}$(G/mM) | 0.1174(8) | 0.1174(8) | 0.1174(8) |
| $V_2$(G/mM) | 0.0530(9) | 0.0543(8) | 0.0542(8) |
| $B_2$(G/mM) | 0.0120(2) | 0.0116(2) | 0.0116(2) |

We also calculate the concentration coefficients of OF parameters obtained by the exact and simplified relations from the SF parameters (Table 2). These coefficients agree with the coefficients of fitted OF parameters within the standard errors. This results support again the consistency between different fitting procedures.

**Fig. 3.** Temperature dependences of the concentration coefficients $W_j$, $V_j$, and $B_j$ for **(a)** $^{15}$N and **(b)** $^{14}$N-pDTEMPONE in water. Errors of coefficients are smaller than symbols.

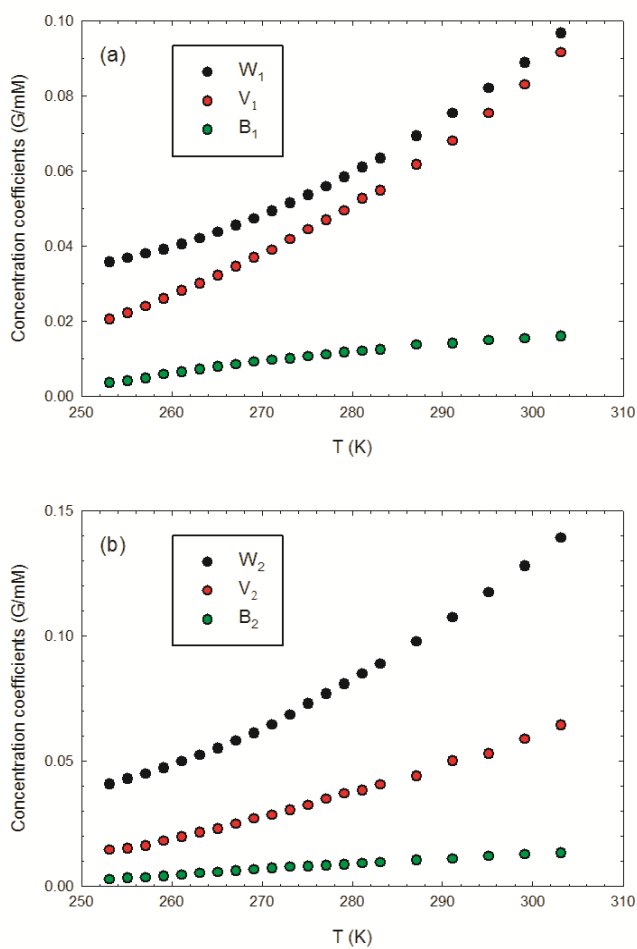

By repeating the OF fitting procedure for EPR spectra and the linear regression method for concentration dependences of fitted parameters at each measured temperature, the concentration coefficients $W_j$, $V_j$, and $B_j$ are obtained as a function of temperature (Fig. 3). It can be seen that all three coefficients have positive values and increase with temperature in the measured temperature range. These concentration coefficients will be analyzed within the continuous diffusion model to obtain diffusion coefficients of both radicals.

**4. Evaluating the effects of HSE and DD interactions on EPR concentration coefficients**

*4.1. Standard relations for the effects of HSE and DD interactions*

In the continuous diffusion model (CDM), the relative motion of radicals A and B is characterized by the relative diffusion coefficient $D_r = D_A + D_B$, where $D_{A,B}$ are the diffusion coefficients of single radicals. The characteristic length is the distance of closest approach of a colliding radical pair $\sigma = r_A + r_B$, where $r_{A,B}$ are the radii of the spheres representing single radicals. The characteristic time of encounter of a diffusing radical pair is given by $\tau_D = \sigma^2 / D_r$, which can be understood as the total time of all reencounters of the radical pair during one encounter [12]. The HSE interaction between radicals has the form $H_{HSE} = \hbar J(r) \vec{S}_A \vec{S}_B$, where the exchange integral $J(r)$ is a strongly decreasing function of the relative distance $r$ between radicals. The HSE interaction can be approximated by the exchange integral having a constant value $J_0$ in the narrow range of relative distances $\sigma \leq r \leq \sigma + \Delta$, where $\Delta$ is a small interaction layer width. This approximation sets another characteristic time $\tau_C = \sigma \Delta / D_r$, which is the contact time or the total time that the radicals spend in the interaction layer during one encounter [12]. The CDM was applied for calculating the effects of DD interaction in the motional narrowing regime $\tau_D \omega_{DD} \ll 1$, where the characteristic time $\tau_D$ is much shorter than inverse of the characteristic DD frequency $\omega_{DD}$ [2,16-18]. In this regime, the contribution of DD interaction on the EPR parameters can be written in terms of the spectral densities of the correlation functions for DD interaction. If, additionally, the characteristic time $\tau_D$ is much longer than the inverse Zeeman frequencies of the radicals ($\tau_D \omega_{A,B} \gg 1$), the contribution of DD interaction is only via correlation functions for its secular part $H_{DD}^{(0)} = \hbar\omega_{DD}(\sigma/r)^3 Y_2^0(\Omega)\left(S_A^+ S_B^- + S_A^- S_B^+ - 4S_A^z S_B^z\right)$. Here, $\Omega$ is the orientation of relative position vector $\vec{r}$ with respect to the applied magnetic field and $\omega_{DD} = \sqrt{\pi/5}(\hbar\gamma_e^2\mu_0)/(4\pi\sigma^3)$ is the characteristic DD frequency. Applying the above approximations on the solution of A and B radicals with respective concentrations $C_A$ and $C_B$, the spin dephasing rate $\gamma_A$, coherence transfer rate $\lambda_A$, and frequency shift $\Delta\omega_A$ of A radicals due to the HSE and DD interactions with B radicals were found to be [2,15-18]:

$$\begin{aligned}
\gamma_A &= k_D C_B \operatorname{Re} p + (C_B \kappa_{DD}^2/k_D)\operatorname{Re} j_\gamma \\
\lambda_A &= k_D C_A \operatorname{Re} p - (C_A \kappa_{DD}^2/k_D)\operatorname{Re} j_\lambda \quad , \\
\Delta\omega_A &= k_D C_B \operatorname{Im} p + (C_B \kappa_{DD}^2/k_D)\operatorname{Im} j_\gamma
\end{aligned} \quad (5)$$

Here, the first and second terms are the contributions from HSE and DD interaction, respectively, $k_D = 4\pi\sigma D_r$ is the rate constant of diffusion encounters, and $\kappa_{DD} = 2\sqrt{\pi}\sigma^3\omega_{DD}$ is the rate constant of DD interaction. The parameters $p$ and $j_{\gamma,\lambda}$ depending on the Zeeman frequency difference $\delta = \omega_A - \omega_B$ satisfy $p(-\delta) = p^*(\delta)$ and $j_{\gamma,\lambda}(-\delta) = j_{\gamma,\lambda}^*(\delta)$. In the $^{15}$N-pDTEMPONE solution, the A and B radicals correspond to any of two subensembles with possible frequency differences $\delta = 0, \pm a$ and concentrations $C_A = C_B = C/2$ [see Eq. (A1)]. Using Eq. (5) for this solution, the concentration coefficients in G/mM units are found to be:

$$\begin{aligned}
W_1 &= \frac{N_A k_D}{2\gamma_e}\operatorname{Re}[p(a)] + \frac{N_A \kappa_{DD}^2}{2\gamma_e k_D}\operatorname{Re}\left[j_\gamma(0) + j_\lambda(0) + j_\gamma(a)\right] \\
V_1 &= \frac{N_A k_D}{2\gamma_e}\operatorname{Re}[p(a)] - \frac{N_A \kappa_{DD}^2}{2\gamma_e k_D}\operatorname{Re}[j_\lambda(a)] \quad , \\
B_1 &= -\frac{N_A k_D}{\gamma_e}\operatorname{Im}[p(a)] - \frac{N_A \kappa_{DD}^2}{\gamma_e k_D}\operatorname{Im}[j_\gamma(a)]
\end{aligned} \quad (6)$$

where $N_A$ is the Avogadro constant and $\gamma_e = g\mu_B/\hbar$ ($g$ is the radical g-factor and $\mu_B$ is the Bohr magneton). The A and B radicals in the $^{14}$N-pDTEMPONE solution correspond to any of three subensembles with possible frequency differences $\delta = 0, \pm a, \pm 2a$ and concentrations $C_A = C_B = C/3$ [see Eq. (A2)]. The concentration coefficients for this solution are:

$$W_2 = \frac{N_A k_D}{3\gamma_e} \text{Re}\left[\frac{4p(a) + 2p(2a)}{3}\right] + \frac{N_A \kappa_{DD}^2}{3\gamma_e k_D} \text{Re}\left[j_\gamma(0) + j_\lambda(0) + \frac{4j_\gamma(a) + 2j_\gamma(2a)}{3}\right]$$

$$V_2 = \frac{N_A k_D}{3\gamma_e} \text{Re}\left[\frac{2p(a) + p(2a)}{3}\right] - \frac{N_A \kappa_{DD}^2}{3\gamma_e k_D} \text{Re}\left[\frac{2j_\lambda(a) + j_\lambda(2a)}{3}\right] \quad , \quad (7)$$

$$B_2 = -\frac{N_A k_D}{3\gamma_e} \text{Im}[p(a) + p(2a)] - \frac{N_A \kappa_{DD}^2}{3\gamma_e k_D} \text{Im}[j_\gamma(a) + j_\gamma(2a)]$$

where $V_2$ is the coherence-transfer coefficient averaged over all pairs of subensembles.

In the standard treatment of HSE effects, a strong exchange ($J_0 \tau_C >> 1$), a narrow interaction layer ($\tau_C << \tau_D$), and a small frequency difference ($|\delta|\tau_D << 1$) are assumed, which gives $\text{Re } p \approx 1/2$ and $\text{Im } p \approx -(\text{sgn } \delta/4)\sqrt{|\delta|\tau_D/2}$ [12]. The DD parameters are related to the spectral densities of correlation functions as $j_\gamma = 4j(0) + j(\delta)$ and $j_\lambda = 2j(0) + 2j(\delta)$, where $j(\omega) = \int_0^\infty du J_{3/2}^2(u)/(u^3 - iu\omega\tau_D)$ is the spectral density of DD interaction in the CDM [2,16]. In the standard treatment, a small frequency difference ($|\delta|\tau_D << 1$) is neglected in the spectral density, which gives $j(\delta) \approx j(0) = 2/15$. Assuming that radical solutions in our case satisfy all above conditions, the concentration coefficients (6-7) take the form:

$$W_1 = \frac{N_A k_D}{4\gamma_e} + \frac{14 N_A \kappa_{DD}^2}{15 \gamma_e k_D}; \quad V_1 = \frac{N_A k_D}{4\gamma_e} - \frac{4 N_A \kappa_{DD}^2}{15 \gamma_e k_D}; \quad B_1 = \frac{N_A k_D}{4\gamma_e} \sqrt{\frac{a\tau_D}{2}}$$

$$W_2 = \frac{N_A k_D}{3\gamma_e} + \frac{38 N_A \kappa_{DD}^2}{45 \gamma_e k_D}; \quad V_2 = \frac{N_A k_D}{6\gamma_e} - \frac{8 N_A \kappa_{DD}^2}{45 \gamma_e k_D}; \quad B_2 = \frac{N_A k_D}{3\gamma_e} \frac{(1+\sqrt{2})}{4} \sqrt{\frac{a\tau_D}{2}} \quad . \quad (8)$$

The relations (8) are referred as the standard relations for the concentration coefficients due to the HSE and DD interactions. We calculated the characteristic time of encounter of a diffusing radical pair $\tau_D$ from all concentration coefficients (8). The averaged experimental values $A$=22.46 G for [15]N-pDTEMPONE and $A$=16.04 G for [14]N-pDTEMPONE were used in the calculation of the hyperfine coupling constant $a = \gamma_e A$. Since the characteristic time is related to the rate constant $k_D$ as $\tau_D = 4\pi\sigma^3/k_D$, the only parameter that should be set in the calculation is the closest distance $\sigma$. For this parameter, we used the value $\sigma = 2r_{vdW}$, where $r_{vdW}$ is the van der Waals radius of pDTEMPONE having the value of 3.5 Å [19]. We get similar values of $\tau_D$ calculated from $W_j$ and $V_j$ for both coefficients and both radicals, which is expected result (Fig. 4). Also, the expected result is that the values of $\tau_D$ calculated from $W_j$ and $V_j$ decrease with the temperature (Fig. 4), because $\tau_D$ is inversely proportional to the relative diffusion coefficient $D_r$, which should increase with temperature. However, the values of $\tau_D$ obtained from the hyperfine splitting coefficients $B_j$ are much higher than the values obtained from $W_j$ and $V_j$ (Fig. 4). Since $B_j \propto \tau_D^{-1/2}$ is predicted by (8), it follows that the standard relations predict much higher $B_j$ than is measured.

The relations (8) imply that the HSE and DDI contributions to the coefficient $W_j$ can be separated by defining $H_j = (b_j W_j + V_j)/(1/j + b_j)$ and $D_j = W_j - H_j$, where $b_1 = 2/7$ for $^{15}$N-pDTEMPONE and $b_2 = 4/19$ for $^{14}$N-pDTEMPONE [20]. It can be easy shown that $H_j = W_j^{HSE}$ and $D_j = W_j^{DD}$ when (8) holds. This separation allows us to calculate $\tau_D$ directly from the relative shift coefficient $\eta_j = B_j / H_j$, without any assumption about the value of $\sigma$. The values of $\tau_D$ were calculated from the relations $\eta_1 = \sqrt{a\tau_D/2}$ ($^{15}$N-pDTEMPONE) and $\eta_2 = (1+\sqrt{2})\sqrt{a\tau_D/32}$ ($^{14}$N-pDTEMPONE), which follow from (8). The calculated $\tau_D$ (Fig. 4) are similar for both radicals, as expected, but they are much lower than $\tau_D$ obtained from $W_j$ and $V_j$, which indicates again that the standard relations (8) predict much higher $B_j$ than is measured. Additionally, the values of $\tau_D$ obtained from $\eta_j$ do not decrease with temperature monotonically, but they show a maximum at about 273 K (Fig. 4). This anomalous behavior has been detected for $^{14}$N-pDTEMPONE in water, suggesting that the total time of reencounters of a radical pair does not follow the hydrodynamic behavior [15].

**Fig. 4.** Temperature dependences of characteristic diffusion time $\tau_D$ calculated from the standard relations (8) and concentration coefficients $W_j$, $V_j$, $B_j$, and $\eta_j$ for **(a)** $^{15}$N- and **(b)** $^{14}$N-pDTEMPONE in water.

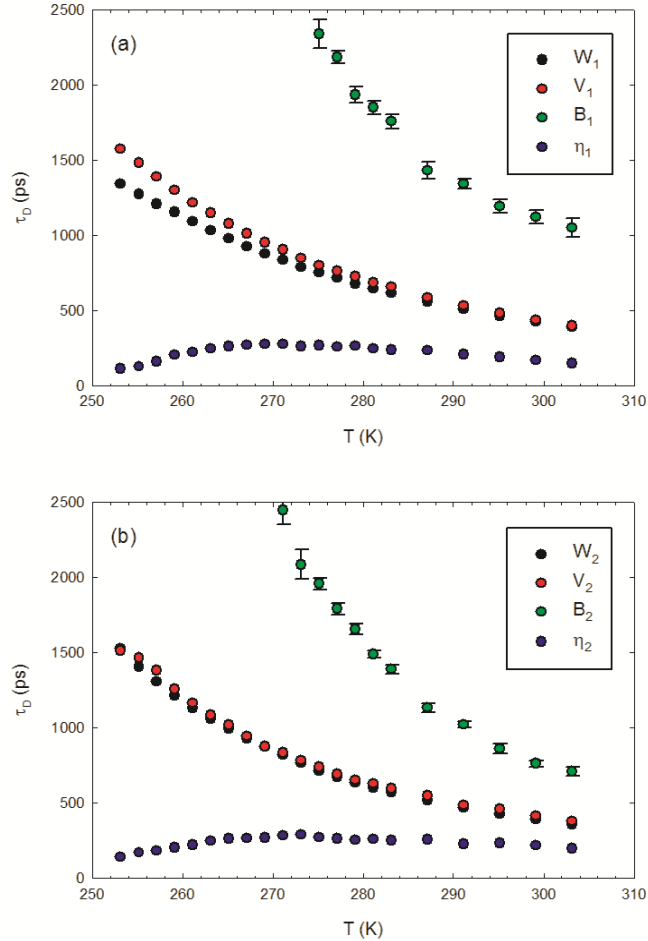

*4.2. Relations from the kinetic equations for the spin density matrix*

In order to resolve the anomalous behavior that arise when using the standard relations for the concentration coefficients, we went beyond these relations by applying the formalism of the kinetic equations for the spin density matrix [1,12,21]. This formalism, which was previously used to calculate the HSE effects on EPR parameters within the CDM, is extended here by including the secular part of DD interaction between radicals.

We considered the kinetic equations for the spin density matrices of a system with A and B radicals with different Zeeman frequencies in external magnetic field [1,12,21]. Relative motion of radicals in the system is described by the CDM and the radicals interact by the HSE interaction and the secular part of DD interaction. We calculated the spin dephasing rate, the coherence transfer rate, and the frequency shift of A radicals produced by the interaction with B radicals (see Appendix B). We assume the strong HSE ($J_0 \tau_C \gg 1$) in the narrow HSE interaction layer ($x_C = \Delta/\sigma \ll 1$), where the contact time is small enough to satisfy $|\delta|\tau_C \ll 1$ and $\omega_{DD}\tau_C \ll 1$. In this case, the parameters in Eq. (5) are determined by the following equations:

$$p = [p_1(\delta) + p_1^*(-\delta)]/2; \; j_\gamma = 2j_1 + j_2; \; j_\lambda = j_1 + 2j_2; \; j_{1,2} = [q_{1,2}(\delta) + q_{1,2}^*(-\delta)]/2;$$
$$p_1(\delta) = \frac{1}{4\pi}\int \left.\frac{\partial T_1(x,\Omega)}{\partial x}\right|_{x=1} d\Omega; \; q_{1,2}(\delta) = \frac{1}{i\beta_D}\int d\Omega Y_2^0(\Omega) \int_1^\infty \frac{T_{1,2}(x,\Omega)dx}{x} \quad (9)$$

where $x = r/\sigma$ ($1 < x < \infty$), $\beta_D = \omega_{DD}\tau_D$, and $T_{1,2}(x,\Omega)$ are the solutions of equations:

$$\frac{1}{x}\frac{\partial^2(xT_1)}{\partial x^2} + \frac{\nabla_\Omega^2 T_1}{x^2} = -i\beta_D Y_2^0(\Omega)\frac{2T_1 + T_2}{x^3}$$
$$\frac{1}{x}\frac{\partial^2(xT_2)}{\partial x^2} + \frac{\nabla_\Omega^2 T_2}{x^2} + i\delta\tau_D T_2 = -i\beta_D Y_2^0(\Omega)\frac{T_1 + 2T_2}{x^3}, \quad (10)$$

where $\nabla_\Omega^2$ is the angular part of Laplacian. The boundary conditions for $x=1$ are $T_1 = T_2$ and $\partial T_1/\partial x = -\partial T_2/\partial x$, while those for $x \to \infty$ are $T_1 \to 1$ and $T_2 \to 0$. Solving Eqs. (10) in the first iteration (FI), the following relations for the parameters (9) are derived (see Appendix C):

$$p_1(\delta) = p_1^*(-\delta) = p = \frac{1+y}{2+y}$$
$$q_1(\delta) = q_1^*(-\delta) = j_1 = \frac{15 + 10y + 2y^2}{9(6 + 4y + y^2)}, \quad (11)$$
$$q_2(\delta) = q_2^*(-\delta) = j_2 = \frac{1}{6 + 4y + y^2}$$

where $y^2 = -i\delta\tau_D$. The formula (11) for the HSE parameter $p$ is the same to the formula in Ref. [12] in the limits of a strong exchange ($J_0\tau_C \gg 1$) and narrow interaction layer ($\tau_C \ll \tau_D$). When the HSE interaction is "switched off", the boundary conditions for $x = 1$ are $\partial T_1/\partial x = \partial T_2/\partial x = 0$. In this case, the DD parameters in FI are $j_\gamma = 4j(0) + j(\delta)$ and $j_\lambda = 2j(0) + 2j(\delta)$, where $j(\delta) = (4+y)/(27 + 27y + 12y^2 + 3y^3)$ is the spectral density. This form of spectral density of DD interaction has already been derived for the CDM with reflecting boundary at $r = \sigma$ [22].

**Fig. 5.** Temperature dependences of characteristic diffusion time $\tau_D$ calculated from the relations (7) and concentration coefficients $W_j$, $V_j$, and $B_j$ for **(a)** $^{15}$N- and **(b)** $^{14}$N-pDTEMPONE in water.

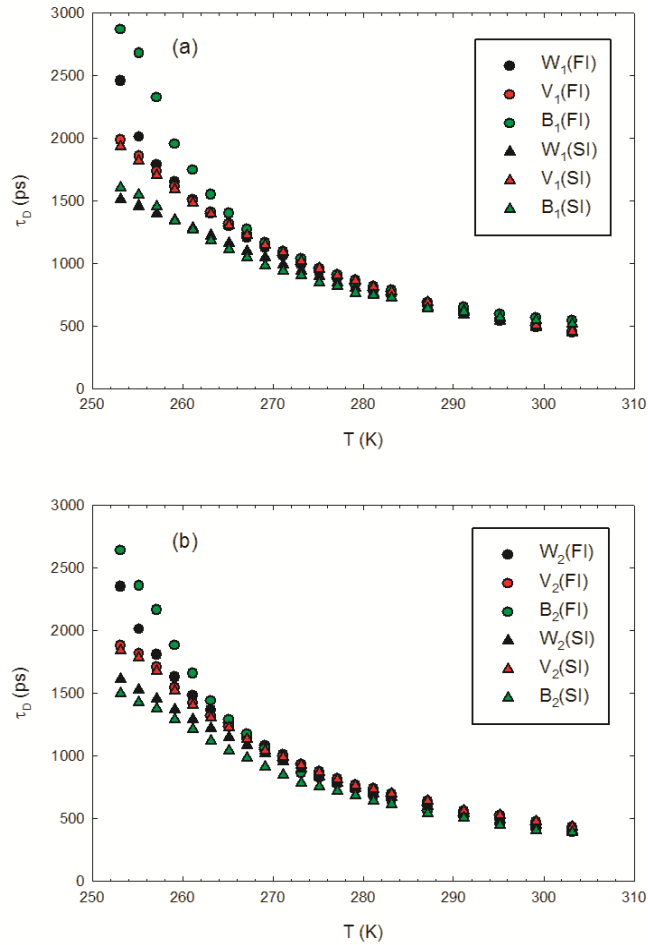

Inserting the parameters (11) into the relations (6) and (7), we calculated $\tau_D$ from all concentration coefficients for $^{15}$N- and $^{14}$N-pDTEMPONE taking the same values of hyperfine constants $a$ and closest distance $\sigma$ as before. The results of the calculation show that all values of $\tau_D$ are now close to each other, especially at higher temperatures (Fig. 5). Also, all values of $\tau_D$ decrease with the temperature without showing anomalous behavior. These results suggest that the previous specific behavior of $\tau_D$ obtained from $B_j$ and $\eta_j$ within the standard treatment is a result of the approximations employed in the standard relations (8).

Applying the second iteration (SI) to the system (10), we calculated the corrections $\Delta p$, $\Delta j_1$, and $\Delta j_2$ of the FI parameters (11). The obtained relations for corrections have complicated forms written down in (C12) and (C13) of Appendix C. By adding these corrections to the parameters (11) and putting them into the relations (6) and (7), we calculated $\tau_D$ in SI from all the concentration coefficients (Fig. 5). It can be seen that all results in the normal state above 273 K are little influenced by applying SI. Additionally, differences between the SI and FI values of $\tau_D$ are the smallest for $\tau_D$ obtained from the coefficients $V_j$ (Fig. 5). Therefore, we propose the values of $\tau_D$ calculated in SI from the coefficients $V_j$ as most reliable values of $\tau_D$ for further calculations.

## 5. Discussion and conclusions

First, we analyze the relative shift coefficients $\eta_j$ whose experimental values exhibit a broad maximum [Fig. 6(a)]. The values of $\tau_D$ that were directly computed from these coefficients using the standard relations (8) exhibit a similar broad maximum (Fig. 4), which was referred as anomalous behavior of $\tau_D$ [15]. On the other hand, the values of $\tau_D$ calculated in the FI and SI from the relations (9-10) exhibit normal behavior, i.e., they decrease with the temperature (Fig. 5). Therefore, it is interesting to calculate coefficients $\eta_j$ form the SI relations for concentration coefficients using the values of $\tau_D$ calculated in SI from $V_j$. We can see that the coefficients $\eta_j$ calculated in this way exhibit a broad maximum too [Fig. 6(a)]. It suggests again that the approximations applied in the standard relations (8) are too crude and produce anomalous behavior of $\tau_D$.

**Fig. 6.(a)** Experimental and calculated values of relative shift coefficient $\eta_j$ for $^{15}$N- and $^{14}$N-pDTEMPONE. **(b)** Deviations of $\tau_D(B_j)$ and $\tau_D(W_j)$ from $\tau_D(V_j)$ at $T=303$ K as a function of the half of closest distance $\sigma/2$ for $^{15}$N- and $^{14}$N-pDTEMPONE.

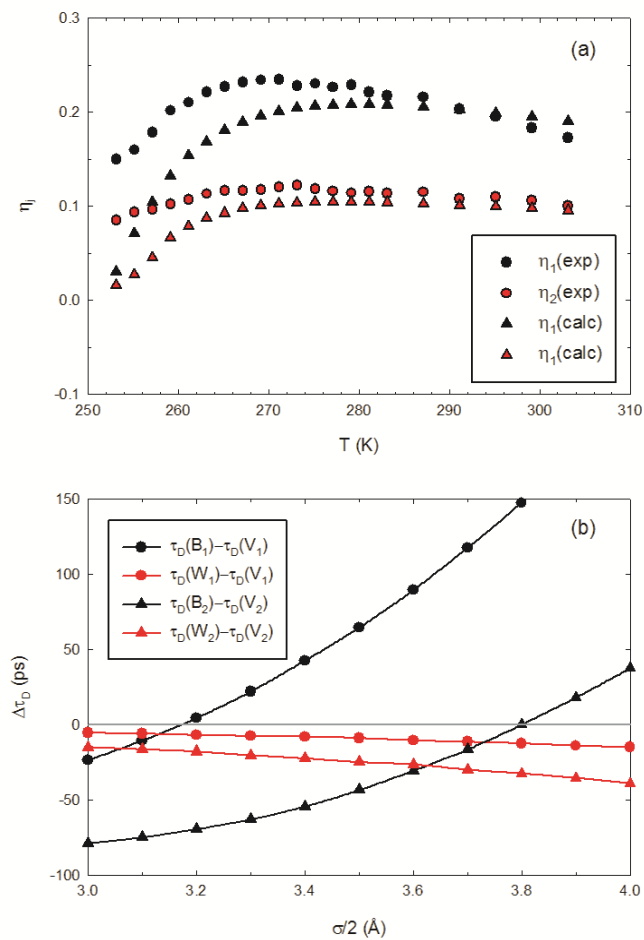

The second point we want to analyze is how a variation of the closest distance $\sigma$, as the only arbitrary parameter in calculations, influences the results. For that reason we calculated $\tau_D$ in SI from the coefficients $W_j$, $V_j$, and $B_j$ at 303 K for various values of $\sigma/2$. The results show that the deviation $\tau_D(B_j) - \tau_D(V_j)$ much strongly depends on $\sigma/2$ than the deviation $\tau_D(W_j) - \tau_D(V_j)$ for both radicals [Fig. 6(b)]. This suggests that the former deviation can be used to judge the best value of $\sigma/2$. The values of $\tau_D(B_j) - \tau_D(V_j)$ are zero at about $\sigma/2 = 3.2$ Å for $^{15}$N-pDTEMPONE and $\sigma/2 = 3.8$ Å for $^{14}$N-pDTEMPONE [Fig. 6(b)]. The mean value of these two $\sigma/2$ are close to the van der Waals radius of 3.5 Å, which justifies our choice of $\sigma/2 = 3.5$ Å for both radicals.

Now, we can check validity of approximations applied in derivation of Eqs. (9-10) for the parameters (5). First, we assumed that the DD interaction affects EPR parameters only through the diffusion-induced modulation of its secular part. This assumption is valid when the parameter $\tau_D \omega_0$ is much larger than one ($\omega_0$ is Zeeman frequency of radical). Using the facts that $\omega_0$ is the X-band frequency ($6 \times 10^{10}$ rad/s) and that the calculated values of $\tau_D$ are between 500 and 2000 ps (Fig. 5), it follows that in our case $\tau_D \omega_0$ is between 30 and 120, which validates our assumption. For validation of approximations made for HSE effects, we can take typical values $J_0 \approx 10^{11}$ rad/s for the HSE integral of radicals at contact distance and $\Delta \approx 0.5$ Å for the HSE interaction layer width [13]. It follows that the assumption of narrow interaction layer is fulfilled in our case because the relative width of interaction layer $x_C = \Delta/\sigma$ has a value of about 0.07. This value of relative width implies that the contact time $\tau_C = x_C \tau_D$ is between 35 and 140 ps. Then it follows that $J_0 \tau_C$ has the values between 3.5 and 14, which agree with the assumed strong exchange limit ($J_0 \tau_C \gg 1$). Using the calculated values for the frequency difference $|\delta| \approx 4 \times 10^8$ rad/s and the DD frequency $\omega_{DD} \approx 8 \times 10^8$ rad/s, we get the values of $|\delta|\tau_C$ from 0.014 to 0.056 and the values of $\omega_{DD}\tau_C$ from 0.028 to 0.112. All these values imply that the contact time of radicals is small enough to satisfy $|\delta|\tau_C \ll 1$ and $\omega_{DD}\tau_C \ll 1$.

**Fig. 7.** Translational diffusion coefficients of $^{15}$N- and $^{14}$N-pDTEMPONE in water obtained from $V_j$ and $\sigma/2 = 3.5$ Å.

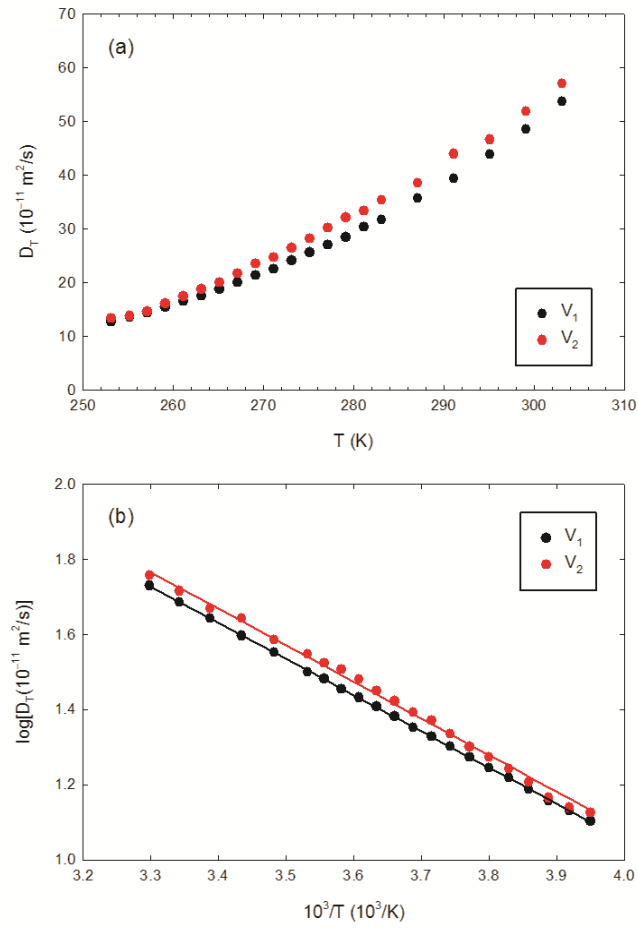

Finally, we calculated the translational diffusion coefficient $D_T = D_r/2$ for both radicals using the relation $D_r = \sigma^2/\tau_D$ and the values of $\tau_D$ calculated in SI from the coefficients $V_j$ (Fig. 5). The obtained results exhibit a good agreement between the values of diffusion coefficients for two radicals with a relative deviation within 10% [Fig. 6(a)]. This result supports our method and shows a benefit of measuring all three concentration coefficients $W_j$, $V_j$, and $B_j$ instead of only $W_j$, as in the line-broadening method [1,3-8]. Temperature dependence of translational diffusion coefficient can be approximated by the Arrhenius law [Fig. 6(b)], where the activation energy has the values 18.43±0.04 and 18.6±0.3 kJ/mol for $^{15}$N-pDTEMPONE and $^{14}$N-pDTEMPONE, respectively.

Usually, temperature dependence of translational diffusion coefficient of a molecule in the normal state of liquid follows the Stokes-Einstien (SE) relation $D_T \propto T/\eta$, where $\eta$ is the viscosity of liquid. The analogous relation for the rotational diffusion coefficient $D_R$ is the Stokes-Einstien-Debye (SED) relation $D_R \propto T/\eta$. It was found that the temperature dependence of viscosity of water in the normal and supercooled states can be well reproduced by a single power law [23]. Using this power law dependence of $\eta$, we can test the SE relation for $D_T$ of $^{15}$N- and $^{14}$N-pDTEMPONE. Deviations from the SE relation can be compared to the deviations from SED relation of the measured $D_R = 1/(6\tau_R)$ for $^{14}$N-pDTEMPONE, where $\tau_R$ is the rotational correlation time [19]. To do this, we define the SE ratio $R_{SE} = D_T \eta / T$ and the SED ratio $R_{SED} = D_R \eta / T$, which should be constant when the SE and SED relations are valid [24]. We calculated the normalized SE ratio $R_{SE}/R_{SE}(303K)$ for $^{15}$N- and $^{14}$N-pDTEMPONE and the normalized SED ratio $R_{SED}/R_{SED}(303K)$ for $^{14}$N-pDTEMPONE [Fig. 8(a)]. The SE ratios increase with decreasing temperature, in the supercooled state, while the SED ratio exhibits a weaker temperature dependence and a slight maximum at about 273 K.

**Fig. 8.(a)** Normalized SE ratio $R_{SE}/R_{SE}(303\ K)$ and normalized SED ratio $R_{SED}/R_{SED}(303\ K)$ for $^{15}$N-pDTEMPONE, $^{14}$N-pDTEMPONE and water molecule in water. **(b)** The ratio between translational and rotational diffusion coefficient $D_T/D_R$ for $^{14}$N-pDTEMPONE and water molecule in water.

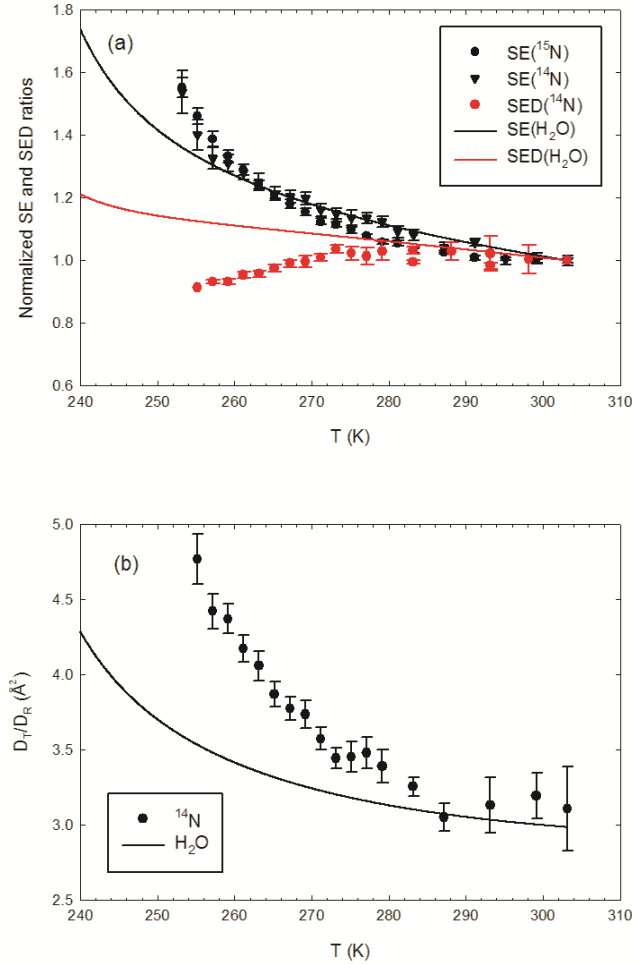

We also calculated the normalized SE and SED ratios for $D_T$ and $D_R$ of water molecules, which were measured by NMR in the normal and supercooled water [25,26]. As can be seen, the SE ratio for water molecule increases with decreasing temperature much strongly than SED ratio [Fig. 8(a)]. It follows that both the pDTEMPONE radical and water molecule show a stronger violation of the SE relation than the SED relation in the supercooled state. This can be further illustrated by calculating the ratio $D_T/D_R$ for $^{14}$N-pDTEMPONE and water molecules [26]. The calculated ratios in both cases increase with lowering the temperature [Fig. 8(b)], indicating decoupling of the translational and rotational motion in the supercooled state [24,26].

In conclusion, we measured EPR spectra for various concentrations of $^{14}$N-pDTEMPONE and $^{15}$N-pDTEMPONE radicals in the normal and supercooled states of water. The EPR parameters of both radicals were calculated by fitting the spectra to the original spectral shape functions derived from the modified Bloch equations. We determined the linear concentration coefficients of the EPR parameters describing the spin dephasing, spin coherence transfer and hyperfine splitting from their concentrations dependences. To evaluate the diffusion coefficients of radicals from the concentration coefficients, we assumed that the radicals interact by the HSE and DD interactions and move according to the CDM. Taking for the closest distance between radicals a value that corresponds to two van der Waals radii of the radical, we applied the standard relations for the concentration coefficients and the relations derived by iterative solving of the kinetic equations for the spin evolution of radical pair. The latter equations were found to reproduce the normal hydrodynamic behavior of diffusion coefficients derived from the hyperfine splitting coefficients, as opposed to the standard relations. Additionally, these equations predict similar values of the diffusion coefficients calculated from all three concentration coefficients for both radicals. The temperature dependences of the calculated diffusion coefficients of radicals were compared to the Stokes-Einstein relation and the temperature dependence of the rotational diffusion coefficient of $^{14}$N-pDTEMPONE. Upon lowering the temperature into the supercooled state, the calculated diffusion coefficients decrease slower than it is predicted by the Stokes-Einstein relation and slower than the rotational diffusion coefficient. Similar effects were detected in NMR studies of the rotational and translational motion of water molecules in supercooled water.

**Acknowledgment**

This work has been fully supported by the Croatian Science Foundation under the project 1108. D.M. wishes to thank Dejana Carić for her assistance in sample preparation.

**Appendix A: Shape of EPR spectra**

$^{15}$N-labeled nitroxide radicals in solution consist of two subensembles having the projections $-1/2$ and $1/2$ of nuclear spin of $^{15}$N [2]. The motion of the transversal magnetizations of subensembles in external magnetic field $B$ and non-saturating microwave field is described by the modified Bloch equations:

$$\begin{aligned}\partial M_1^-/\partial t &= -i(\omega_1 - \omega)M_1^- - \gamma_1 M_1^- + \lambda M_2^- - i\omega_m M_0 \\ \partial M_2^-/\partial t &= -i(\omega_2 - \omega)M_2^- - \gamma_2 M_2^- + \lambda M_1^- - i\omega_m M_0\end{aligned} \quad (A1)$$

Here, $\omega$ is microwave frequency, $\gamma_k$ is spin dephasing rate of subensemble $k$, $\lambda$ is coherence transfer rate, $\omega_m$ is the nutation frequency, and $M_0$ is equilibrium value of longitudinal subensemble magnetization. The frequencies of subenembles are $\omega_1 = \omega_0 - a/2$ and $\omega_2 = \omega_0 + a/2$, where $\omega_0$ is Zeeman frequency of radical and $a$ is the hyperfine coupling constant. For $^{14}$N-labeled nitroxide radicals, there are three subensembles having projections $-1$, $0$, and $1$ of nuclear spin of $^{14}$N and the modified Bloch equations have the form:

$$\begin{aligned}\partial M_1^-/\partial t &= -i(\omega_1 - \omega)M_1^- - \gamma_1 M_1^- + \lambda M_2^- + \lambda M_3^- - i\omega_m M_0 \\ \partial M_2^-/\partial t &= -i(\omega_2 - \omega)M_2^- - \gamma_2 M_2^- + \lambda M_1^- + \lambda M_3^- - i\omega_m M_0 \\ \partial M_3^-/\partial t &= -i(\omega_3 - \omega)M_3^- - \gamma_3 M_3^- + \lambda M_1^- + \lambda M_2^- - i\omega_m M_0\end{aligned} \quad (A2)$$

where $\omega_1 = \omega_0 - a$, $\omega_2 = \omega_0$ and $\omega_3 = \omega_0 + a$. By solving (A1) and (A2) for the stationary condition $\partial M_k^-/\partial t = 0$, total magnetization turns out to be:

$$M^- = \sum_{k=1}^{2I+1} M_k^- = -i\omega_m M_0 \frac{g(\omega)}{1 - \lambda g(\omega)}; \quad g(\omega) = \sum_{k=1}^{2I+1} \frac{1}{\gamma_k + \lambda + i(\omega_k - \omega)}. \quad (A3)$$

This relation implies that the absorption EPR signal has the form:

$$R(\omega) = -\operatorname{Im} M^- = \omega_m M_0 \operatorname{Re}\left[\frac{g(\omega)}{1-\lambda g(\omega)}\right]. \tag{A4}$$

Since the Zeeman frequency is $\omega_0 = \gamma_e B$, where $\gamma_e = g\mu_B/\hbar$ ($g$ is radical g-factor and $\mu_B$ is Bohr magneton), the frequency variables in (A3) can be replaced by field variables using relations: $\omega = \gamma_e B_0, \lambda = \gamma_e \Lambda, \gamma_i = \gamma_e \Gamma_i$, and $a = \gamma_e A$. This leads to the absorption EPR signal (1), where it is taken into account that the outer lines of $^{14}$N-labeled radical move relative to the central one by a small second-order hyperfine shift $\Delta A$. Since the absorption EPR signal (1) can be written as the sum function (2), this function can be used to fit EPR spectra. Then, the obtained parameters $f_k$ and $z'_k$ should be transformed into the original ones $J_0$, $\Lambda$, and $z_k$ from Eqs. (1). Defining the auxiliary variables $X_k \equiv (Z_k, Z'_k, F_k)$ related to main variables $x_k \equiv (z_k, z'_k, f_k)$ as:

$$X_1 = (x_1 + x_2)/2,\ X_2 = (x_1 - x_2)/2 \tag{A5a}$$

for $^{15}$N-labeled radical and

$$X_1 = (x_1 + x_2 + x_3)/3,\ X_2 = (x_1 - x_3)/2,\ X_3 = (x_1 - 2x_2 + x_3)/6 \tag{A5b}$$

for $^{14}$N-labeled radical, the original parameters can be calculated using consecutively the relations:

$$J_0 = \operatorname{Re}(F_1),\ \Lambda = -\operatorname{Re}(F_2 Z'_2)/J_0,\ Z_1 = Z'_1,\ Z_2^2 = Z_2'^2 - \Lambda^2 \tag{A6a}$$

for $^{15}$N-labeled radical and the relations:

$$\begin{aligned}
&J_0 = \operatorname{Re}(F_1),\ \Lambda = -\operatorname{Re}(F_2 Z'_2/3 + F_3 Z'_3)/J_0,\\
&Z_1 = Z'_1,\ Z_3 = (W - Q/W)/2,\ Z_2^2 = -3(Z_3^2 + Q),\\
&W^3 = R + \sqrt{R^2 + Q^3},\ Q = -Z_3'^2 - Z_2'^2/3 + \Lambda^2,\ R = Z_3'^3 - Z_2'^2 Z'_3 - \Lambda^3;
\end{aligned} \tag{A6b}$$

for $^{14}$N-labeled radical. It follows from $Z_1 = Z'_1$ that $\Gamma = \Gamma'$, where $\Gamma$ is the average linewidth of hyperfine lines. By neglecting the second-order hyperfine shift and differences between linewidths of hyperfine lines in $Z_2$, $Z'_2$, $Z_3$, and $Z'_3$, the other two simplified relations from Eqs.(3) can be obtained from (A6).

**Appendix B: EPR parameters from the kinetic equation for the density matrix**

The eigenstates of Zeeman Hamiltonian of A and B spins $H_0 = \hbar(\omega_A S^z_A + \omega_B S^z_B)$ are: $|1\rangle = |++\rangle$, $|2\rangle = |+-\rangle$, $|3\rangle = |-+\rangle$ and $|4\rangle = |--\rangle$, where the first and second symbols denote the sign of $S^z_A$ and $S^z_B$ eigenvalues, respectively. Considering only the secular term of DD interaction, the interaction between A and B spins $V = H_{HSE} + H^{(0)}_{DD}$ has non-zero matrix elements in the basis of $H_0$ eigenstates: $V_{22} = V_{33} = -V_{11} = -V_{44} = \hbar\Omega_D/2$ and $V_{23} = V_{32} = \hbar\Omega_N$, where $\Omega_D = -J(r)/2 + 2J^{(0)}_{DD}(\vec{r})$, $\Omega_N = J(r)/2 + J^{(0)}_{DD}(\vec{r})$, and $J^{(0)}_{DD}(\vec{r}) = \omega_{DD}(\sigma/r)^3 Y^0_2(\Omega)$. The equation of motion for the density matrix of A spins is:

$$d\sigma_A/dt = -i\omega_A [S^z_A, \sigma_A] - C_B \text{Tr}_B(\hat{P}\sigma_A \otimes \sigma_B). \tag{B1}$$

Here, $\hat{P} = -\int d\vec{r}\,\hat{\psi}(\vec{r})$ is the impact (super)operator, where $\hat{\psi}(\vec{r}) = \hat{\Omega}(\vec{r})\hat{T}(\vec{r})$ and $\vec{r}$ is the relative position vector between A and B spins. The interaction operator $\hat{\Omega}(\vec{r})$ has elements:

$$\hat{\Omega}_{ij,mn} = -(i/\hbar)(V_{im}\delta_{jn} - V_{nj}\delta_{im}) \tag{B2}$$

and the correlation operator $\hat{T}(\vec{r},t)$ satisfies $\hat{\rho}(r,t) = \hat{T}(r,t)\sigma_A(t) \otimes \sigma_B(t)$, where $\hat{\rho}(r,t)$ is the partial density matrix of isolated AB pairs with the relative position $\vec{r}$. In the CDM, the stationary correlation operator $\hat{T}(r) = \lim_{t\to\infty} \hat{T}(r,t)$ satisfies the equation $-D_r\nabla^2\hat{T} - [\hat{Q},\hat{T}] = \hat{\Omega}\hat{T}$ and the boundary conditions $\hat{T}(\vec{r}) \to \hat{I}$ for $r \to \infty$ and $\partial\hat{T}(\vec{r})/\partial r = 0$ for $r = \sigma$. In the $H_0$ basis, the Zeeman interaction operator $\hat{Q}$ has only diagonal elements $\hat{Q}_{ij,ij} = -i\Delta\omega_{ij}$, where $\Delta\omega_{ij} = (H_{0ii} - H_{0jj})/\hbar$ and the equation for operator element $\hat{T}_{ij,kl}$ is:

$$-D_r \nabla^2 \hat{T}_{ij,kl} + i(\Delta\omega_{ij} - \Delta\omega_{kl})\hat{T}_{ij,kl} = \sum_{m,n} \hat{\Omega}_{ij,mn} \hat{T}_{mn,kl} = \hat{\psi}_{ij,kl}. \tag{B3}$$

The equation of motion (B1) for the off-diagonal element of density matrix is:

$$\dot{\sigma}_{+-}^A = -i\omega_A \sigma_{+-}^A + C_B \int d\vec{r} \sum_{k,l} (\hat{\psi}_{13,kl} + \hat{\psi}_{24,kl})(\sigma_A \otimes \sigma_B)_{kl}. \tag{B4}$$

As follows from (B2), the equations (B3) determining the operators in (B4) are:

$$\begin{aligned}
\hat{\psi}_{13,kl} &= -D_r \nabla^2 \hat{T}_{13,kl} + i(\omega_A - \Delta\omega_{kl})\hat{T}_{13,kl} = i\Omega_D \hat{T}_{13,kl} + i\Omega_N \hat{T}_{12,kl} \\
\hat{\psi}_{12,kl} &= -D_r \nabla^2 \hat{T}_{12,kl} + i(\omega_B - \Delta\omega_{kl})\hat{T}_{12,kl} = i\Omega_D \hat{T}_{12,kl} + i\Omega_N \hat{T}_{13,kl} \\
\hat{\psi}_{24,kl} &= -D_r \nabla^2 \hat{T}_{24,kl} + i(\omega_A - \Delta\omega_{kl})\hat{T}_{24,kl} = -i\Omega_D \hat{T}_{24,kl} - i\Omega_N \hat{T}_{34,kl} \\
\hat{\psi}_{34,kl} &= -D_r \nabla^2 \hat{T}_{34,kl} + i(\omega_B - \Delta\omega_{kl})\hat{T}_{34,kl} = -i\Omega_D \hat{T}_{34,kl} - i\Omega_N \hat{T}_{24,kl}
\end{aligned} \tag{B5}$$

Since the non-zero components of $\hat{T}$ are diagonal ones and those coupled with them, we get four pairs of equations from (B5). They couple the pairs $(\hat{T}_{13,13}, \hat{T}_{12,13})$, $(\hat{T}_{13,12}, \hat{T}_{12,12})$, $(\hat{T}_{24,24}, \hat{T}_{34,24})$, and $(\hat{T}_{24,34}, \hat{T}_{34,34})$ determining respectively nonzero terms $\hat{\psi}_{13,13}$, $\hat{\psi}_{13,12}$, $\hat{\psi}_{24,24}$, and $\hat{\psi}_{24,34}$ in (B4). Using the replacements $\hat{T}_{13,13} \to T_1$, $\hat{T}_{12,13} \to T_2$, $\hat{\psi}_{13,13} \to \psi_1(\delta)$, and $\hat{\psi}_{12,13} \to \psi_2(\delta)$, the first pair of equations becomes:

$$\begin{aligned}
\psi_1(\delta) &= -D_r \nabla^2 T_1 = -iJ(r)(T_1 - T_2)/2 + i\omega_{DD} Y_2^0(\Omega)(\sigma/r)^3 (2T_1 + T_2) \\
\psi_2(\delta) &= -D_r \nabla^2 T_2 - i\delta T_2 = iJ(r)(T_1 - T_2)/2 + i\omega_{DD} Y_2^0(\Omega)(\sigma/r)^3 (T_1 + 2T_2)
\end{aligned}, \tag{B6}$$

while the quantities from other three equations are given by:

$$\begin{aligned}
\hat{T}_{13,12} &= T_2(-\delta); \ \hat{T}_{12,12} = T_1(-\delta); \ \hat{\psi}_{13,12} = \psi_2(-\delta) \\
\hat{T}_{24,24} &= T_1^*(-\delta); \ \hat{T}_{34,24} = T_2^*(-\delta); \ \hat{\psi}_{24,24} = \psi_1^*(-\delta). \\
\hat{T}_{24,34} &= T_2^*; \ \hat{T}_{34,34} = T_1^*; \ \hat{\psi}_{24,34} = \psi_2^*
\end{aligned} \tag{B7}$$

The above consideration implies that (B4) can be written as:

$$\dot{\sigma}_{+-}^A = -i\omega_A \sigma_{+-}^A - C_B \frac{P_1(\delta) + P_1^*(-\delta)}{2} \sigma_{+-}^A - C_B \frac{P_2(-\delta) + P_2^*(\delta)}{2} \sigma_{+-}^B, \tag{B8}$$

where the impact operator components are $P_{1,2}(\delta) = -\int d\vec{r}\, \psi_{1,2}(\delta)$. In deriving (B8) from (B4), we used relations $(\sigma_A \otimes \sigma_B)_{13} = \sigma_{+-}^A \sigma_{++}^B$, $(\sigma_A \otimes \sigma_B)_{12} = \sigma_{++}^A \sigma_{+-}^B$, $(\sigma_A \otimes \sigma_B)_{24} = \sigma_{+-}^A \sigma_{--}^B$, and $(\sigma_A \otimes \sigma_B)_{34} = \sigma_{--}^A \sigma_{+-}^B$ with assumption $\sigma_{++}^{A,B} \approx \sigma_{--}^{A,B} \approx 1/2$. Since $M_A^- \propto C_A \sigma_{+-}^A$, the spin dephasing rate, the coherence transfer rate, and the frequency shift are given by:

$$\begin{aligned}
\gamma_A &= (C_B/2)\operatorname{Re}\left[P_1(\delta) + P_1^*(-\delta)\right] \\
\lambda_A &= -(C_A/2)\operatorname{Re}\left[P_2(-\delta) + P_2^*(\delta)\right] \\
\Delta\omega_A &= (C_B/2)\operatorname{Im}\left[P_1(\delta) + P_1^*(-\delta)\right]
\end{aligned} \qquad (B9)$$

We introduce relative distance variable $x = r/\sigma$ and assume that $J(x) = J_0$ in a narrow interaction layer $1 \leq x \leq 1 + x_C$ ($x_C = \Delta/\sigma \ll 1$). The equalities $T_{1,2}(B) = T_{1,2}^I(B)$ and $(\partial T_{1,2}/\partial x)_B = (\partial T_{1,2}^I/\partial x)_B$ hold at the interaction zone boundary $B \equiv 1 + x_C, \Omega$, where $T_{1,2}^I$ are the solutions of (B6) within the interaction layer. We integrate (B6) over radial variable within the interaction layer assuming $T_{1,2}^I(x,\Omega) \approx T_{1,2}^I(B)$ and taking into account the equalities at $B$ and $(\partial T_{1,2}^I/\partial x)_{1,\Omega} = 0$. Taking $x_C \ll 1$ in the integrals, $T_{1,2}$ at $B \approx 1, \Omega$ satisfies:

$$\begin{aligned}
(\partial T_1/\partial x)_B &= i\frac{J_0 \tau_C}{2}(T_1 - T_2)_B - i\omega_{DD}\tau_C Y_2^0(\Omega)(2T_1 + T_2)_B \\
(\partial T_2/\partial x)_B + i\delta\tau_C T_2(B) &= i\frac{J_0 \tau_C}{2}(T_2 - T_1)_B - i\omega_{DD}\tau_C Y_2^0(\Omega)(T_1 + 2T_2)_B
\end{aligned} \qquad (B10)$$

Additional integrating of the left side of (B10) over angular variables gives the following relations for HSE impact operator components:

$$\begin{aligned}
4\pi P_1^{HSE}(\delta)/k_D &= \int d\Omega (\partial T_1/\partial x)_B \\
4\pi P_2^{HSE}(\delta)/k_D &= \int d\Omega (\partial T_2/\partial x)_B + i\delta\tau_C \int d\Omega\, T_2(B)
\end{aligned} \qquad (B11)$$

Assuming $|\delta|\tau_C \ll 1$, $\omega_{DD}\tau_C \ll 1$, and strong exchange $J_0\tau_C \gg 1$, the boundary conditions (B10) are $T_1(1,\Omega) = T_2(1,\Omega)$ and $(\partial T_1/\partial x)_{1,\Omega} = -(\partial T_2/\partial x)_{1,\Omega}$, while the HSE impact operator components are:

$$P_1^{HSE}(\delta) = -P_2^{HSE}(\delta) = k_D p_1(\delta)$$
$$p_1(\delta) = (1/4\pi)\int d\Omega (\partial T_1/\partial x)_{1,\Omega} \quad . \tag{B12}$$

Only DD interaction exists for $x > 1+x_C$ and the equations (B6) after multiplying them by $-\tau_D$ take the form (10). By integrating (10) and neglecting interaction layer width ($x_C \to 0$), the DD impact operator components are:

$$P_1^{DD}(\delta) = (\kappa_{DD}^2/k_D)[2q_1(\delta) + q_2(\delta)]$$
$$P_2^{DD}(\delta) = (\kappa_{DD}^2/k_D)[q_1(\delta) + 2q_2(\delta)] \quad . \tag{B13}$$
$$q_{1,2}(\delta) = -(i/\beta_D)\int d\Omega Y_2^0(\Omega)\int_1^\infty dx T_{1,2}(x,\Omega)/x$$

It follows from (B9), (B12), and (B13) that the spin dephasing rate, the coherence transfer rate, and the frequency shift are given by Eq. (5), with the parameters defined by Eqs. (9-10).

**Appendix C: Iterative calculation of EPR parameters**

We solve the equations (10) iteratively by taking $T_1 = T_1(x \to \infty) = 1$ and $T_2 = T_2(x \to \infty) = 0$ on the right-hand side. By defining $t_1 = xT_1$ and $t_2 = xT_2$, the equations take the form:

$$\frac{\partial^2 t_1}{\partial x^2} + \frac{\nabla_\Omega^2 t_1}{x^2} = -2i\beta_D \frac{Y_2^0(\Omega)}{x^2}$$
$$\frac{\partial^2 t_2}{\partial x^2} + \frac{\nabla_\Omega^2 t_2}{x^2} - y^2 t_2 = -i\beta_D \frac{Y_2^0(\Omega)}{x^2} \quad . \tag{C1}$$

where $y^2 = -i\delta\tau_D$. Since generally we can write $t_1(x,\Omega) = \sum_{l=0}^{\infty} u_l(x) Y_{2l}^0(\Omega)$, and $t_2(x,\Omega) = \sum_{l=0}^{\infty} v_l(x) Y_{2l}^0(\Omega)$, it follows from (C1) and $\nabla_\Omega^2 Y_{2l}^0(\Omega) = -2l(2l+1) Y_{2l}^0(\Omega)$ that radial functions satisfy the equations:

$$u_l''(x) - \frac{2l(2l+1)}{x^2} u_l(x) = -2i\beta_D \frac{1}{x^2} \delta_{l,1}$$
$$v_l''(x) - \frac{2l(2l+1)}{x^2} v_l(x) - y^2 v_l(x) = -i\beta_D \frac{1}{x^2} \delta_{l,1}$$
(C2)

The solutions of (C2) for $l=0$ and $l=1$, which satisfy the boundary conditions $T_1(x \to \infty) \to 1$ and $T_2(x \to \infty) \to 0$ can be written as:

$$u_0(x) = c_0 + (4\pi)^{1/2} x; \quad v_0(x) = d_0 e^{-y(x-1)}$$
$$u_1(x) = i\beta_D \left( \frac{c_1}{x^2} + \frac{1}{3} \right); \quad v_1(x) = i\beta_D \left[ d_1 e^{-y(x-1)} y_2\left(\frac{1}{yx}\right) + \frac{1}{(yx)^2} \right]$$
(C3)

where $y_2$ is the Bessel polynomial of degree 2. The constants $c_{0,1}$ and $d_{0,1}$ have to be obtained from the boundary conditions $T_1(1,\Omega) = T_2(1,\Omega)$ and $(\partial T_1/\partial x)_{1,\Omega} = -(\partial T_2/\partial x)_{1,\Omega}$, which imply $u_l(1) = v_l(1)$ and $u_l'(1) - u_l(1) = v_l(1) - v_l'(1)$. By solving these equations for $l=0$ and $l=1$, we get:

$$c_0(y) = -(4\pi)^{1/2} \frac{1+y}{2+y}; \quad d_0(y) = (4\pi)^{1/2} \frac{1}{2+y}$$
$$c_1(y) = -\frac{1}{3} \frac{y^2 + 2y + 3}{y^2 + 4y + 6}; \quad d_1(y) = \frac{2}{3} \frac{y-3}{y^2 + 4y + 6}$$
(C4)

Inserting the constants (C4) into the parameters of interest (9), which have the form:

$$p_1(\delta) = \frac{1}{4\pi} \int d\Omega (\partial T_1 / \partial x)_{1,\Omega} = -\frac{c_0}{(4\pi)^{1/2}}$$
$$q_1(\delta) = \int_1^\infty \frac{u_1(x)}{i\beta_D} \frac{dx}{x^2} = \frac{c_1}{3} + \frac{1}{3}; \quad q_2(\delta) = \int_1^\infty \frac{v_1(x)}{i\beta_D} \frac{dx}{x^2} = d_1 \frac{1+y}{y^2} + \frac{1}{3y^2}$$
(C5)

we get the relations (11) for parameters in the first iteration (FI). Since the solutions of (C2) that satisfy the boundary conditions for $l>1$ are $u_l(x)=0$ and $v_l(x)=0$, it follows that the solutions of (C1) are:

$$t_1(x,\Omega) = u_0(x)Y_0^0(\Omega) + u_1(x)Y_2^0(\Omega)$$
$$t_2(x,\Omega) = v_0(x)Y_0^0(\Omega) + v_1(x)Y_2^0(\Omega)$$
(C6)

where the radial functions are given by (C3) with the constants (C4). In the second iteration (SI), we replace $T_{1,2}$ on the right-hand side of equations (10) by the extra terms $t_1/x-1$ and $t_2/x$ obtained from the FI solutions (C6). This leads to the equations:

$$\frac{\partial^2 \Delta t_1}{\partial x^2} + \frac{\nabla_\Omega^2 \Delta t_1}{x^2} = -i\beta_D Y_2^0 (2f_{u0} + f_{v0}) + (\beta_D Y_2^0)^2 (2f_{u1} + f_{v1})$$

$$\frac{\partial^2 \Delta t_2}{\partial x^2} + \frac{\nabla_\Omega^2 \Delta t_2}{x^2} - y^2 t_2 = -i\beta_D Y_2^0 (f_{u0} + 2f_{v0}) + (\beta_D Y_2^0)^2 (f_{u1} + 2f_{v1})$$
(C7)

for the corrections $\Delta t_{1,2} = x\Delta T_{1,2}$ to the FI solutions, where

$$f_{u0}(x) = \frac{c_0}{(4\pi)^{1/2} x^3}; \quad f_{v0}(x) = \frac{d_0 e^{-y(x-1)}}{(4\pi)^{1/2} x^3}$$

$$f_{u1}(x) = \frac{c_1}{x^5} + \frac{1}{3x^3}; \quad f_{v1}(x) = \frac{d_1 e^{-y(x-1)}}{x^3} y_2\left(\frac{1}{yx}\right) + \frac{1}{(yx)^2 x^3}$$
(C8)

Inserting $\Delta t_1(x,\Omega) = \sum_{l=0}^{\infty} \Delta u_l(x) Y_{2l}^0(\Omega)$ and $\Delta t_2(x,\Omega) = \sum_{l=0}^{\infty} \Delta v_l(x) Y_{2l}^0(\Omega)$ into (C7), the radial functions for $l=0$ are fond to satisfy:

$$\Delta u_0''(x) = \frac{\beta_D^2}{(4\pi)^{1/2}} [2f_{u1}(x) + f_{v1}(x)]$$

$$\Delta v_0''(x) - y^2 \Delta v_0(x) = \frac{\beta_D^2}{(4\pi)^{1/2}} [f_{u1}(x) + 2f_{v1}(x)]$$
(C9)

while those for $l=1$ satisfy:

$$\Delta u_1''(x) - \frac{6\Delta u_1(x)}{x^2} = -i\beta_D \left[ 2f_{u0}(x) + f_{v0}(x) \right] + \beta_D^2 k_{2,2} \left[ 2f_{u1}(x) + f_{v1}(x) \right]$$
$$\Delta v_1''(x) - \frac{6\Delta v_1(x)}{x^2} - y^2 \Delta v_1(x) = -i\beta_D \left[ f_{u0}(x) + 2f_{v0}(x) \right] + \beta_D^2 k_{2,2} \left[ f_{u1}(x) + 2f_{v1}(x) \right]$$
(C10)

where $k_{2,2} = \int d\Omega \left[ Y_2^0(\Omega) \right]^3$. Finding solutions of (C9) and (C10) that satisfy boundary conditions $\Delta T_{1,2}(x \to \infty) \to 0$, $\Delta T_1(1,\Omega) = \Delta T_2(1,\Omega)$, and $(\partial \Delta T_1 / \partial x)_{1,\Omega} = -(\partial \Delta T_2 / \partial x)_{1,\Omega}$, we calculate the corrections of parameters (9):

$$\Delta p_1(\delta) = (4\pi)^{-1/2} \left[ \Delta u_0'(1) - \Delta u_0(1) \right]$$
$$\Delta q_1(\delta) = \int_1^\infty \frac{\Delta u_1(x)}{i\beta_D} \frac{dx}{x^2}; \quad \Delta q_2(\delta) = \int_1^\infty \frac{\Delta v_1(x)}{i\beta_D} \frac{dx}{x^2}$$
(C11)

The result can be written as:

$$\Delta p_1(\delta) = \Delta p_1^*(-\delta) = \Delta p = -\frac{\beta_D^2}{4\pi} \frac{a_1(y) + e^y y^2 a_2(y)}{72(2+y)(6+4y+y^2)}$$
$$\Delta q_1(\delta) = \Delta q_1^*(-\delta) = \Delta j_1 = \frac{b_{11}(y) + e^y y^2 b_{12}(y)}{144(2+y)(3+y)(6+4y+y^2)},$$
$$\Delta q_2(\delta) = \Delta q_2^*(-\delta) = \Delta j_2 = -\frac{b_{21}(y) + e^y y^2 b_{22}(y)}{8(2+y)(3+y)(6+4y+y^2)}$$
(C12)

where

$$a_1(y) = 18 + 294y + 161y^2 + 21y^3 - y^4 - y^5$$
$$a_2(y) = (126 + 84y + 9y^2 - 2y^3 - y^4)\text{Ei}(-y)$$
$$b_{11}(y) = -108 - 648y - 588y^2 - 188y^3 - 9y^4 + 4y^5 + y^6$$
$$b_{12}(y) = 6y^2(1+y)\left[\text{Chi}(y) - \text{Shi}(y)\right] + (-216 - 216y - 72y^2 + 5y^3 + y^4)\text{Ei}(-y)$$
$$b_{21}(y) = -6 + 28y + 21y^2 + 4y^3 + y^4$$
$$b_{22}(y) = (18 + 18y + 5y^2 + y^3)\text{Ei}(-y) + 3(1+y)\left[\ln(-1/y) - \ln(-y) + 2\ln(y)\right]$$
(C13)

Here, Ei, Chi, and Shi denote exponential, hyperbolic cosine and hyperbolic sine integrals, respectively.